\documentclass[aps,prl,reprint,superscriptaddress,showpacs]{revtex4-1}
\usepackage{natbib}
\usepackage{mathptmx} 
\usepackage{pstricks}
\usepackage{array}
\usepackage{tabularx}
\usepackage{graphicx}
\usepackage{color}
\usepackage{psfrag}
\usepackage{amsmath}
\usepackage{amssymb}
\usepackage{psfrag}
\usepackage{textcomp}
\usepackage{longtable}
\usepackage{multirow}
\usepackage{braket}
\usepackage[mathscr]{euscript}
\usepackage{dcolumn}
\usepackage{bm}
\usepackage{hyperref}
\hypersetup{letterpaper, colorlinks=true, urlcolor=blue, citecolor=blue, linkcolor=blue, breaklinks=true}
\usepackage{breakurl}

\newcommand{\msu}{Department of Physics and Astronomy, Michigan State University, East Lansing, Michigan 48824, USA}
\newcommand{\msuchem}{Department of Chemistry, Michigan State University, East Lansing, Michigan 48824, USA
}
\newcommand{\nscl}{National Superconducting Cyclotron Laboratory, Michigan State University, East Lansing, Michigan 48824, USA}
\newcommand{\augustana}{Department of Physics and Astronomy, Augustana College, Rock Island, Illinois 61201, USA}
\newcommand{\wabash}{Department of Physics, Wabash College, Crawfordsville, Indiana 47933, USA}
\newcommand{\hope}{Department of Physics, Hope College, Holland, Michigan 49423, USA}
\newcommand{\cmu}{Department of Physics, Central Michigan University, Mt. Pleasant, Michigan 48859, USA}
\newcommand{\iusb}{Department of Physics and Astronomy, Indiana University at South Bend, South Bend, Indiana 46634, USA}
\newcommand{\westmont}{Department of Physics, Westmont College, Santa Barbara, California 93108, USA}
\newcommand{\triumf}{TRIUMF, 4004 Wesbrook Mall, Vancouver, British Columbia V6T 2A3, Canada}
\newcommand{\cord}{Department of Physics, Concordia College, Moorhead, Minnesota 56562, USA}

\newcommand{\ganil}{GANIL, CEA/DSM-CNRS/IN2P3, Bvd Henri Becquerel, 14076 Caen, France}

\newcommand{\meas}[2]{$#1~\textrm{#2}$}
\newcommand{\nuc}[2]{$^{#1}\textrm{#2}$}
\newcommand{\nucn}[3]{$^{#1}\textrm{#2}^{#3}$}

\newcommand{\figref}[1]{Fig.~\ref{#1}}
\newcommand{\eqnref}[1]{Eq.~\eqref{#1}}

\newcommand{\executeiffilenewer}[3]{%
\if{\Filemodnewest[1]{#1}{#2} == #2}
{\immediate\write18{#3}}\fi%
}

\begin{document}
\title{Exploring the Low-{Z} Shore of the Island of Inversion at {N} = 19}

\author{G. Christian}
\email{chris402@msu.edu}
\thanks{Present address: \triumf}
\affiliation{\msu}
\affiliation{\nscl}

\author{N. Frank}
\affiliation {\augustana}

\author{S. Ash}
\affiliation {\augustana}

\author{T. Baumann}
\affiliation {\nscl}

\author{D. Bazin}
\affiliation {\nscl}

\author{J. Brown}
\affiliation{\wabash}

\author{P. A. DeYoung}
\affiliation{\hope}

\author{J. E. Finck}
\affiliation{\cmu}

\author{A. Gade}
\affiliation{\msu}
\affiliation{\nscl}

\author{G. F. Grinyer}
\thanks{Present address: \ganil}
\affiliation{\nscl}

\author{A. Grovom}
\affiliation{\westmont}

\author{J. D. Hinnefeld}
\affiliation{\iusb}

\author{E. M. Lunderberg}
\affiliation{\hope}

\author{B. Luther}
\affiliation{\cord}

\author{M. Mosby}
\affiliation{\nscl}
\affiliation{\msuchem}

\author{S. Mosby}
\affiliation{\msu}
\affiliation{\nscl}

\author{T. Nagi}
\affiliation{\hope}

\author{G. F. Peaslee}
\affiliation{\hope}

\author{W. F. Rogers}
\affiliation{\westmont}

\author{J. K. Smith}
\affiliation{\msu}
\affiliation{\nscl}

\author{J. Snyder}
\affiliation{\msu}
\affiliation{\nscl}

\author{A. Spyrou}
\affiliation {\msu}
\affiliation {\nscl}

\author{M. J. Strongman}
\affiliation{\msu}
\affiliation{\nscl}

\author{M. Thoennessen}
\affiliation{\msu}
\affiliation{\nscl}

\author{M. Warren}
\affiliation {\augustana}

\author{D. Weisshaar}
\affiliation{\nscl}

\author{A. Wersal}
\affiliation{\nscl}

\date{\today}

\begin{abstract}

The technique of invariant mass spectroscopy has been used to measure, for the first time, the ground state energy of neutron-unbound $^{28}\textrm{F},$ determined to be a resonance in the $^{27}\textrm{F} + n$ continuum at $2\underline{2}0 (\underline{5}0)$ keV. States in $^{28}\textrm{F}$ were populated by the reactions of a $62$ MeV/u $^{29}\textrm{Ne}$ beam impinging on a $288$ $\textrm{mg/cm}^2$ beryllium target. The measured $^{28}\textrm{F}$ ground state energy is in good agreement with USDA/USDB shell model predictions, indicating that $pf$ shell intruder configurations play only a small role in the ground state structure of $^{28}\textrm{F}$ and establishing a low-$Z$ boundary of the island of inversion for $N=19$ isotones.

\end{abstract}


\maketitle

A hallmark of the nuclear shell model is its reproduction of large energy gaps at nucleon numbers 2, 8, 20, 28, 50, 82, and 126.  Although well established in stable nuclei, these magic numbers begin to disappear for  nuclei far from stability.  For example, it has been known for over 30 years that the large shell gap at $N = 20$ diminishes for neutron-rich nuclei \cite{PhysRevC.12.644, PhysRevC.18.2342, PhysRevC.19.164, GuillemaudMueller198437}.  The change in shell structure around $N=20$ is now known to be a result of the tensor force, which is strongly attractive for the spin flip pairs $\pi d _{5/2}$-$\nu d_{3/2}$ and strongly repulsive for the pairs $\pi d _{5/2}$-$\nu f_{7/2}$ \cite{PhysRevLett.87.082502, PhysRevLett.95.232502, PhysRevLett.97.162501}.  For nuclei in the region of $N \sim 20$ and $Z \lesssim 13,$ the reduced $N=20$ gap allows $pf$ shell intruder configurations, in the form of multi-particle, multi-hole ($np$-$nh$ or $n \hslash \omega$) cross shell excitations, to compete with standard $sd$ only configurations if the gain in correlation energy is on the same order as the size of the shell gap \cite{PhysRevC.70.044307, Poves1987311, springerlink:10.1140/epja/i2001-10243-7}.  This has led to the establishment of the ``island of inversion''---a region of nuclei near $N=20$ for which the intruder configuration is dominant in the ground state.

The island of inversion was originally thought to only include those nuclei with $10 \leq Z \leq 12$ and $20 \leq N \leq 22$ \cite{PhysRevC.41.1147}. In more recent years, it has become clear that the island extends further, and much experimental effort has been put forth to determine its boundaries \cite{Gade2008161}.  On the low-$N$ (western) and high-$Z$ (northern) sides of the island---both in the direction of increasing stability---it is generally agreed that ground state intruder components fade away for $Z \geq 13$ and $N \leq 18$. Ground state observables for nuclei lying outside these limits are well described by $sd$ shell model calculations, such as those utilizing the USDA/USDB effective interactions \cite{PhysRevC.74.034315}.  Heading away from stability, the role of intruder configurations in nuclear ground states becomes less clear.  On the high-$N$ (eastern) side, there are strong indications that ground state intruder dominance persists in heavier isotopes of Mg, Na, and Ne. For example, recent measurements of $2p$ knockout cross sections from \nuc{38}{Si} to \nuc{36}{Mg} have indicated a $0 \hslash \omega$ ground state occupation of only $38(8)\%$ in \nuc{36}{Mg} \cite{PhysRevLett.99.072502}, extending the region of inversion to at least $N=24$ for the Mg isotopic chain.

Until now, the low-$Z$ (southern) side of the island has been almost completely unexplored. A measurement of bound excited states in \nuc{27}{F}, which lies on the island's western border at $N=18$, has hinted at $pf$ shell contributions to its \emph{excited} state structure \cite{Elekes200417}, but mass measurements \cite{Jurado200743} indicate that the \nuc{27}{F} ground state is primarily $sd$ shell.  For the heavier $(N \geq 19)$ fluorine isotopes, lying within the island's western boundary, no direct experimental information is available.  All that is known are dripline systematics, e.g. as outlined in \cite{Thoennessen2003C61}. The fluorine isotopic chain is also the only area in which the intruder structure of $Z < 10,~ N \geq 19$ nuclei can be examined in any detail since the neutron dripline for all lighter elements is located at $N \leq 16$ \cite{Tarasov199764, Sakurai1999180, Langevin198571, PhysRevC.41.937, PhysRevC.53.647}. This means that the $N = 19$ isotones of oxygen and below are unbound with respect to three or more neutrons, making their study extremely difficult, if not impossible.

In this letter, we report on the first experimental investigation of the low-$Z$ border of the island of inversion for $N \geq 19$ nuclei.  This is done via a measurement of the ground state of neutron-unbound \nuc{28}{F}, performed with the technique of invariant mass spectroscopy. Our results are then compared to binding energy (mass defect) predictions of the USDA/USDB shell model.  In particular, investigation of the $N = 19$ binding energy systematics provides strong evidence that only the isotones at $Z = 10$, $11$, and $12$ are located within the island.  In addition to mapping the island of inversion, the evolution of intruder structure in neutron-rich fluorine isotopes is relevant to the abrupt shift in the neutron dripline observed between oxygen (ending at $N=16$) and fluorine (ending at $N \geq 22$) \cite{PhysRevC.64.011301}.

The experiment was performed at the National Superconducting Cyclotron Laboratory (NSCL) at Michigan State University, using a primary beam of \nucn{48}{Ca}{20+} accelerated to \meas{140}{MeV/u} in the coupled K500-K1200 cyclotrons \cite{marti:64}. The \nuc{48}{Ca} beam was fragmented in a $1316~\textrm{mg/cm}^2$ beryllium production target, and fragmentation products were selected in the A1900 fragment separator \cite{Morrissey200390}, with the third and fourth segments set to a rigidity of \meas{3.47}{Tm} and a momentum acceptance of $3.9 \%$ to optimize the transmission of \nuc{29}{Ne} fragments.  The secondary beam was passed through a pair of plastic timing scintillators, the first located at the A1900 focal plane and the second \meas{44.3}{cm} upstream of a secondary reaction target.  Additionally, the beam was sent through a pair of position sensitive Cathode Readout Drift Chambers (CRDCs) and a focusing quadrupole triplet.  The desired \nuc{29}{Ne} composed approximately $2 \%$ of the beam and could be fully separated from other components using time of flight (ToF) and energy loss measurements.  The incoming rate of \nuc{29}{Ne} beam particles was approximately $70~\textrm{s}^{-1},$ and the median energy of the \nuc{29}{Ne} was \meas{62}{MeV/u.}

The secondary reaction target was beryllium with a thickness of $288~\textrm{mg/cm}^2.$  States in neutron-unbound \nuc{28}{F} were populated by one-proton knockout and decayed by neutron emission to $^{27}\textrm{F} + n$ with a timescale on the order of $10^{-21}~\textrm{s}.$  The decays of the unbound states potentially feed bound excited states in \nuc{27}{F}, necessitating the measurement of neutrons, charged fragments, and $\gamma$-rays.  The gammas were detected using the CAESAR CsI(Na) array \cite{Weisshaar2010615}, which surrounded the target and provided an in-beam detection efficiency of $\sim 30 \%$ for \meas{1}{MeV} gammas.  Charged fragments were deflected $43^{\circ}$ by the Sweeper magnet \cite{1439869} and passed through a pair of CRDCs, an ionization chamber, and thin $(0.5~\textrm{cm})$ and thick $(15~\textrm{cm})$ plastic scintillators.  Neutrons were detected in the Modular Neutron Array (MoNA) \cite{Baumann2005517}, whose front face was located \meas{658}{cm} downstream of the target at $0^{\circ}.$  MoNA was partially shadowed by the entrance flange of the Sweeper vacuum box, resulting in a neutron angular acceptance of roughly $\pm 6.0^{\circ}$ in the dispersive $(x)$ plane and $\pm 2.5^{\circ}$ in the non-dispersive $(y)$ plane.

\begin{figure}
    \centering
\begingroup
  \makeatletter
  \providecommand\color[2][]{%
    \errmessage{(Inkscape) Color is used for the text in Inkscape, but the package 'color.sty' is not loaded}
    \renewcommand\color[2][]{}%
  }
  \providecommand\transparent[1]{%
    \errmessage{(Inkscape) Transparency is used (non-zero) for the text in Inkscape, but the package 'transparent.sty' is not loaded}
    \renewcommand\transparent[1]{}%
  }
  \providecommand\rotatebox[2]{#2}
  \ifx\svgwidth\undefined
    \setlength{\unitlength}{239.74880371pt}
  \else
    \setlength{\unitlength}{\svgwidth}
  \fi
  \global\let\svgwidth\undefined
  \makeatother
  \begin{picture}(1,0.65456225)%
    \put(0,0){\includegraphics[width=\unitlength]{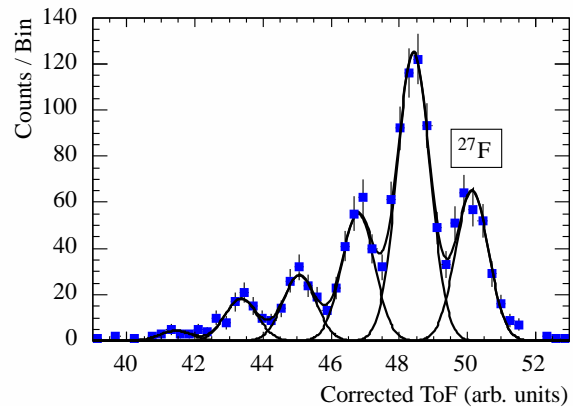}}%
    \put(0.90124227,0.00585242){\color[rgb]{0,0,0}\makebox(0,0)[rb]{\smash{Corrected ToF (arb. units)}}}%
    \put(0.20227978,0.05511948){\color[rgb]{0,0,0}\makebox(0,0)[b]{\smash{40}}}%
    \put(0.3098016,0.05487142){\color[rgb]{0,0,0}\makebox(0,0)[b]{\smash{42}}}%
    \put(0.41745777,0.05487142){\color[rgb]{0,0,0}\makebox(0,0)[b]{\smash{44}}}%
    \put(0.52500024,0.05511948){\color[rgb]{0,0,0}\makebox(0,0)[b]{\smash{46}}}%
    \put(0.73809422,0.05511948){\color[rgb]{0,0,0}\makebox(0,0)[b]{\smash{50}}}%
    \put(0.84561612,0.05511948){\color[rgb]{0,0,0}\makebox(0,0)[b]{\smash{52}}}%
    \put(0.12564437,0.09523817){\color[rgb]{0,0,0}\makebox(0,0)[rb]{\smash{0}}}%
    \put(0.12564437,0.16754856){\color[rgb]{0,0,0}\makebox(0,0)[rb]{\smash{20}}}%
    \put(0.12564437,0.24162267){\color[rgb]{0,0,0}\makebox(0,0)[rb]{\smash{40}}}%
    \put(0.12564437,0.31393316){\color[rgb]{0,0,0}\makebox(0,0)[rb]{\smash{60}}}%
    \put(0.12564437,0.38624364){\color[rgb]{0,0,0}\makebox(0,0)[rb]{\smash{80}}}%
    \put(0.12564437,0.46031775){\color[rgb]{0,0,0}\makebox(0,0)[rb]{\smash{100}}}%
    \put(0.05288127,0.61120405){\color[rgb]{0,0,0}\rotatebox{90}{\makebox(0,0)[rb]{\smash{Counts / Bin}}}}%
    \put(0.12564437,0.53086451){\color[rgb]{0,0,0}\makebox(0,0)[rb]{\smash{120}}}%
    \put(0.6306189,0.05511948){\color[rgb]{0,0,0}\makebox(0,0)[b]{\smash{48}}}%
    \put(0.72585443,0.39015243){\color[rgb]{0,0,0}\makebox(0,0)[lb]{\smash{$^{27}\textrm{F}$}}}%
    \put(0.12564437,0.60320275){\color[rgb]{0,0,0}\makebox(0,0)[rb]{\smash{140}}}%
  \end{picture}%
\endgroup

	\caption{(Color online) Corrected ToF for fluorine isotopes produced from \nuc{29}{Ne}.}
	\label{fig:pid}
\end{figure}

The charged particle measurements allowed for event-by-event isotope selection.  Elements were selected using cuts on both $\Delta E$-ToF and $\Delta E$-$E$, with $\Delta E$ taken from the ion chamber signal and $E$ taken from the total charge collected in the thick scintillator.  For a given element, isotopes were separated by correcting their ToF for the various paths taken through the Sweeper. This amounts to removing correlations between ToF and a variety of other measured parameters, the most important being the dispersive position and angle exiting the Sweeper (up to fourth order), non-dispersive position exiting the magnet, and dispersive position of the incoming beam.  A plot of the corrected ToF for fluorine elements is shown in \figref{fig:pid}, with \nuc{27}{F} indicated.

The decay energy, $E_d,$ of the breakup of unbound states was calculated using invariant mass analysis:
\begin{equation}
\label{eq:edecay}
	E_d =
		\sqrt{m_f^2 + m_n^2 + 2 \left(E_f E_n  - p_f p_n \cos{\theta} \right)}  - m_f - m_n,
\end{equation}

where $m_{f} (m_n),$ $E_f (E_n),$ and $p_f (p_n)$ refer to the mass, energy, and momentum of the charged fragment (neutron), respectively, and $\theta$ is the opening angle between the two decay products.  The neutron input to \eqnref{eq:edecay} was calculated from ToF and position measurements in MoNA using linear kinematics, while the charged fragment input was reconstructed from measurements of the post-Sweeper emittance and the $x$ position of the beam on target \cite{Frank20071478}.

\begin{figure}
    \centering
\begingroup
  \makeatletter
  \providecommand\color[2][]{%
    \errmessage{(Inkscape) Color is used for the text in Inkscape, but the package 'color.sty' is not loaded}
    \renewcommand\color[2][]{}%
  }
  \providecommand\transparent[1]{%
    \errmessage{(Inkscape) Transparency is used (non-zero) for the text in Inkscape, but the package 'transparent.sty' is not loaded}
    \renewcommand\transparent[1]{}%
  }
  \providecommand\rotatebox[2]{#2}
  \ifx\svgwidth\undefined
    \setlength{\unitlength}{239.74880371pt}
  \else
    \setlength{\unitlength}{\svgwidth}
  \fi
  \global\let\svgwidth\undefined
  \makeatother
  \begin{picture}(1,0.65083577)%
    \put(0,0){\includegraphics[width=\unitlength]{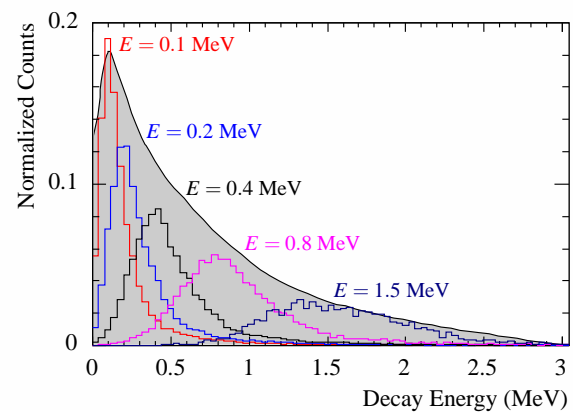}}%
    \put(0.14946573,0.05585987){\color[rgb]{0,0,0}\makebox(0,0)[b]{\smash{0}}}%
    \put(0.27324288,0.05590122){\color[rgb]{0,0,0}\makebox(0,0)[b]{\smash{0.5}}}%
    \put(0.39643614,0.05561185){\color[rgb]{0,0,0}\makebox(0,0)[b]{\smash{1}}}%
    \put(0.51929877,0.05590122){\color[rgb]{0,0,0}\makebox(0,0)[b]{\smash{1.5}}}%
    \put(0.64174627,0.05561185){\color[rgb]{0,0,0}\makebox(0,0)[b]{\smash{2}}}%
    \put(0.76546142,0.05590122){\color[rgb]{0,0,0}\makebox(0,0)[b]{\smash{2.5}}}%
    \put(0.88889757,0.05585987){\color[rgb]{0,0,0}\makebox(0,0)[b]{\smash{3}}}%
    \put(0.05762457,0.61101932){\color[rgb]{0,0,0}\rotatebox{90}{\makebox(0,0)[rb]{\smash{Normalized Counts}}}}%
    \put(0.1256444,0.09439247){\color[rgb]{0,0,0}\makebox(0,0)[rb]{\smash{0}}}%
    \put(0.12747352,0.34840483){\color[rgb]{0,0,0}\makebox(0,0)[rb]{\smash{0.1}}}%
    \put(0.90080055,0.00663448){\color[rgb]{0,0,0}\makebox(0,0)[rb]{\smash{Decay Energy (MeV)}}}%
    \put(0.19183338,0.56657142){\color[rgb]{1,0,0}\makebox(0,0)[lb]{\smash{\footnotesize{$E = 0.1$ MeV}}}}%
    \put(0.53045775,0.17902887){\color[rgb]{0,0,0.50196078}\makebox(0,0)[lb]{\smash{\footnotesize{$E = 1.5$ MeV}}}}%
    \put(0.29765349,0.34041942){\color[rgb]{0,0,0}\makebox(0,0)[lb]{\smash{\footnotesize{$E = 0.4$ MeV}}}}%
    \put(0.23504329,0.43037476){\color[rgb]{0,0,1}\makebox(0,0)[lb]{\smash{\footnotesize{$E = 0.2$ MeV}}}}%
    \put(0.39024609,0.25310973){\color[rgb]{1,0,1}\makebox(0,0)[lb]{\smash{\footnotesize{$E = 0.8$ MeV}}}}%
    \put(0.12747352,0.60239652){\color[rgb]{0,0,0}\makebox(0,0)[rb]{\smash{0.2}}}%
  \end{picture}%
\endgroup

	\caption{(Color online) Simulated resolution and acceptance of the experimental setup.  Each colored histogram was generated by simulating a \nuc{28}{F} breakup at the indicated relative energy, then folding in detector resolution and acceptance cuts. The shaded curve was generated by simulating a \nuc{28}{F} breakup with the relative energy uniformly distributed from $0$--$3~\textrm{MeV}$ and folding in acceptance and resolution. The colored histograms are all normalized to a total area of unity, and the shaded curve was arbitrarily scaled to fit within the same panel.}
	\label{fig:res}
\end{figure}

Resonant states were modeled by a Breit-Wigner lineshape with energy dependent width, derived from $R$-Matrix theory \cite{RevModPhys.30.257}.  Free parameters are the central resonance energy $E_0$, the resonance width $\Gamma_0,$  and the relative contribution to the overall decay spectrum.  The orbital angular momentum was fixed at $\ell = 2$ assuming the emission of a $0d_{3/2}$ neutron.  This assumption may be incorrect if intruder components are significant in \nuc{28}{F}; however, separate analyses of the data using $\ell = 1$ and $\ell = 3$ resonances give results that do not differ significantly from the $\ell = 2$ case.  We also investigated the possibility of $\ell = 0$ decay by modelling the data as an $s$-wave \cite{Blanchon2007} with the scattering length, $a_s,$ freely varying.

Smearing from experimental resolution and acceptance was accounted for in a Monte Carlo simulation of the experimental setup, including all relevant detector resolutions and acceptance cuts. The simulation included a realistic incoming beam profile and used the Goldhaber model \cite{Goldhaber1974306} to describe the one-proton knockout reaction populating \nuc{28}{F}. A demonstration of the simulated overall resolution and acceptance functions is shown in \figref{fig:res}.  To determine optimal fit parameters, large Monte Carlo data sets ($\sim 3$ million events) were generated and compared to the experimental data using an unbinned maximum likelihood technique \cite{Schmidt1993547}.


\begin{figure}
    \centering
\begingroup
  \makeatletter
  \providecommand\color[2][]{%
    \errmessage{(Inkscape) Color is used for the text in Inkscape, but the package 'color.sty' is not loaded}
    \renewcommand\color[2][]{}%
  }
  \providecommand\transparent[1]{%
    \errmessage{(Inkscape) Transparency is used (non-zero) for the text in Inkscape, but the package 'transparent.sty' is not loaded}
    \renewcommand\transparent[1]{}%
  }
  \providecommand\rotatebox[2]{#2}
  \ifx\svgwidth\undefined
    \setlength{\unitlength}{239.74919667pt}
  \else
    \setlength{\unitlength}{\svgwidth}
  \fi
  \global\let\svgwidth\undefined
  \makeatother
  \begin{picture}(1,0.66832853)%
    \put(0,0){\includegraphics[width=\unitlength]{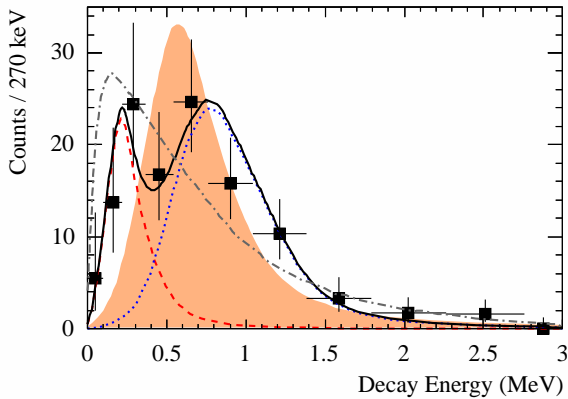}}%
    \put(0.90124151,0.02214542){\color[rgb]{0,0,0}\makebox(0,0)[rb]{\smash{Decay Energy (MeV)}}}%
    \put(0.14990659,0.07137097){\color[rgb]{0,0,0}\makebox(0,0)[b]{\smash{0}}}%
    \put(0.27544742,0.07137097){\color[rgb]{0,0,0}\makebox(0,0)[b]{\smash{0.5}}}%
    \put(0.39864071,0.07137097){\color[rgb]{0,0,0}\makebox(0,0)[b]{\smash{1}}}%
    \put(0.52326696,0.07137097){\color[rgb]{0,0,0}\makebox(0,0)[b]{\smash{1.5}}}%
    \put(0.64924181,0.07137097){\color[rgb]{0,0,0}\makebox(0,0)[b]{\smash{2}}}%
    \put(0.77472064,0.07137097){\color[rgb]{0,0,0}\makebox(0,0)[b]{\smash{2.5}}}%
    \put(0.89992056,0.07137097){\color[rgb]{0,0,0}\makebox(0,0)[b]{\smash{3}}}%
    \put(0.12874779,0.11162015){\color[rgb]{0,0,0}\makebox(0,0)[rb]{\smash{0}}}%
    \put(0.13057689,0.25639745){\color[rgb]{0,0,0}\makebox(0,0)[rb]{\smash{10}}}%
    \put(0.12887177,0.40283727){\color[rgb]{0,0,0}\makebox(0,0)[rb]{\smash{20}}}%
    \put(0.12975016,0.54755384){\color[rgb]{0,0,0}\makebox(0,0)[rb]{\smash{30}}}%
    \put(0.0517788,0.62823762){\color[rgb]{0,0,0}\rotatebox{90}{\makebox(0,0)[rb]{\smash{Counts / 270 keV}}}}%
  \end{picture}%
\endgroup

	\caption{(Color online) Measured decay energy spectrum (including smearing from experimental resolution and acceptance) for $^{27}\textrm{F} + n$ coincidences. The filled squares with error bars are the measured data, and the dashed red and dotted blue curves represent the $220$ keV and $810$ keV simulation results, respectively.  The solid black curve is the sum of the two resonances, with the ratio of $220$ keV resonance to the total area being $28\%.$ The filled orange curve is a simulation of a single resonance at $590$ keV, and the grey dot-dashed curve is the best fit of a single $s$-wave ($a_s = -0.05$ fm).}
	\label{fig:f28edecay}
\end{figure}

The measured decay energy of \nuc{28}{F} into $^{27}\textrm{F} + n$ (with no unfolding of resolution or acceptance) is shown in \figref{fig:f28edecay}. Comparison with \figref{fig:res} reveals that the measured data are strongly distorted by resolution and acceptance. In particular, the width of the measured data is almost entirely due to experimental resolution, and the shape of the data above $\sim 0.8$ MeV is dominated by the limited acceptance at higher relative energies.  No coincident gamma events were recorded in CAESAR, so the observed transitions are assumed to feed the ground state of \nuc{27}{F} (around $30$ CAESAR counts would be expected in the case of $100 \%$ branching to an exited state).   States in \nuc{28}{F} were populated in a direct one-proton knockout reaction from \nuc{29}{Ne}, so non-resonant contributions to the data are not expected.  An attempt to fit the data with a single Breit Wigner resonance is shown as the shaded orange curve in the figure. In this fit, the width was limited to $\Gamma_0 \leq 1$ MeV, beyond which the function saturates and lineshapes become indistinguishable.  The optimal one-resonance fit parameters are $E_0 = 590$ keV and $\Gamma_0 = 1$ MeV.  The $\Gamma_0$ value is two orders of magnitude larger than the single particle prediction, already suggesting that a single resonance cannot properly describe the data. Furthermore, a visual comparison of simulation and data indicates a poor fit, with the simulated curve being significantly narrower than the data despite its unphysically large $\Gamma_0$ value.  The best-fit of a single $s$-wave ($a_s = -0.05$ fm) is shown as the grey dot-dashed curve and is clearly not consistent with the data. 

The poor fit of a single resonance suggests that multiple resonances are present in the data, and a good agreement between simulation and data is achieved by modelling with two independent Breit Wigner resonances, with the width of each resonance fixed at its approximate single-particle value.  Given this model, the best fit is obtained with the lower resonance at $2\underline{2}0 (\underline{5}0)$ keV ($\Gamma_0 \equiv 10$ keV) and the upper resonance at $810$ keV ($\Gamma_0 \equiv 100$ keV), as shown in \figref{fig:f28edecay}. The relative contributions of the lower and upper resonances to the total area are $28 \%$ and $72\%$, respectively. The stronger population of an excited state is somewhat surprising; however, as discussed below, contributions from multiple unresolved excited state resonances may be present in the data. Thus it is not possible to draw any definite conclusions from the relative areas of the two resonances. To examine the statistical significance of the two-resonance versus one-resonance hypotheses, we calculate the likelihood ratio ($-2 \ln[L_1/L_2],$ where $L_1$ and $L_2$ are the respective likelihoods of the one-resonance and two-resonance hypotheses) to be $22.6,$ providing a moderate level of support for the two-resonance model.

As mentioned, it is possible that more than two resonances are present in the data, but the low statistics and limited experimental resolution prevent the inclusion of additional resonances into the fit with any sort of certainty regarding their location. However, attempts to fit the data with three or more resonances demonstrate that the location of the ground state peak is insensitive to the presence of multiple resonances. It should be noted that the placement of the ground state peak at $220$ keV is primarily dictated by the shape of the spectrum below $\sim 300$ keV and not by the dip at around $400$ keV. In particular, modelling with a strong resonance at lower decay energies ($\lesssim 150$ keV) results in a sharp peak that is not consistent with the gradual rise from zero seen in the data. Likewise, placing the lowest resonance too high results in a model that significantly under-predicts the number of events below $\sim 300$ keV.

\begin{figure}
    \centering
\begingroup
  \makeatletter
  \providecommand\color[2][]{%
    \errmessage{(Inkscape) Color is used for the text in Inkscape, but the package 'color.sty' is not loaded}
    \renewcommand\color[2][]{}%
  }
  \providecommand\transparent[1]{%
    \errmessage{(Inkscape) Transparency is used (non-zero) for the text in Inkscape, but the package 'transparent.sty' is not loaded}
    \renewcommand\transparent[1]{}%
  }
  \providecommand\rotatebox[2]{#2}
  \ifx\svgwidth\undefined
    \setlength{\unitlength}{239.74880371pt}
  \else
    \setlength{\unitlength}{\svgwidth}
  \fi
  \global\let\svgwidth\undefined
  \makeatother
  \begin{picture}(1,0.64985839)%
    \put(0,0){\includegraphics[width=\unitlength]{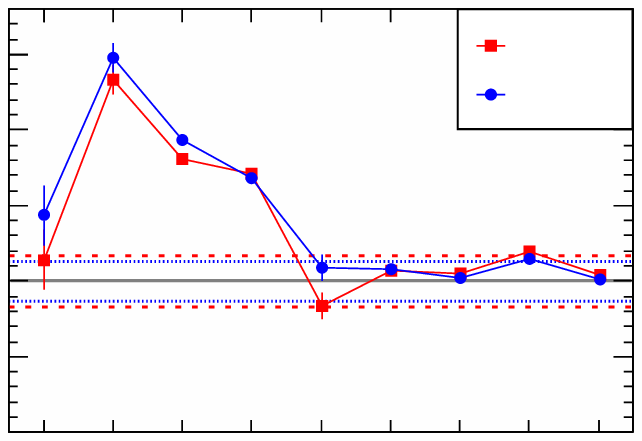}}%
    \put(0.90095153,0.00367852){\color[rgb]{0,0,0}\makebox(0,0)[rb]{\smash{$Z$}}}%
    \put(0.19227578,0.05590115){\color[rgb]{0,0,0}\makebox(0,0)[b]{\smash{9}}}%
    \put(0.2742691,0.05585981){\color[rgb]{0,0,0}\makebox(0,0)[b]{\smash{10}}}%
    \put(0.35807604,0.05561174){\color[rgb]{0,0,0}\makebox(0,0)[b]{\smash{11}}}%
    \put(0.43923745,0.05561174){\color[rgb]{0,0,0}\makebox(0,0)[b]{\smash{12}}}%
    \put(0.52345432,0.05585981){\color[rgb]{0,0,0}\makebox(0,0)[b]{\smash{13}}}%
    \put(0.60685821,0.05561174){\color[rgb]{0,0,0}\makebox(0,0)[b]{\smash{14}}}%
    \put(0.68941991,0.05585981){\color[rgb]{0,0,0}\makebox(0,0)[b]{\smash{15}}}%
    \put(0.77260165,0.05585981){\color[rgb]{0,0,0}\makebox(0,0)[b]{\smash{16}}}%
    \put(0.85719561,0.05590115){\color[rgb]{0,0,0}\makebox(0,0)[b]{\smash{17}}}%
    \put(0.05337725,0.61199925){\color[rgb]{0,0,0}\rotatebox{90}{\makebox(0,0)[rb]{\smash{$\textit{BE}_\textit{exp} - \textit{BE}_\textit{th}$ (MeV)}}}}%
    \put(0.12747351,0.09523675){\color[rgb]{0,0,0}\makebox(0,0)[rb]{\smash{-1}}}%
    \put(0.12628513,0.18694753){\color[rgb]{0,0,0}\makebox(0,0)[rb]{\smash{-0.5}}}%
    \put(0.1256444,0.27689458){\color[rgb]{0,0,0}\makebox(0,0)[rb]{\smash{0}}}%
    \put(0.12628513,0.36860535){\color[rgb]{0,0,0}\makebox(0,0)[rb]{\smash{0.5}}}%
    \put(0.12747351,0.4585524){\color[rgb]{0,0,0}\makebox(0,0)[rb]{\smash{1}}}%
    \put(0.12628513,0.54849945){\color[rgb]{0,0,0}\makebox(0,0)[rb]{\smash{1.5}}}%
    \put(0.75387055,0.55897703){\color[rgb]{0,0,0}\makebox(0,0)[lb]{\smash{USDA}}}%
    \put(0.75387055,0.50043398){\color[rgb]{0,0,0}\makebox(0,0)[lb]{\smash{USDB}}}%
  \end{picture}%
\endgroup

	\caption{(Color online) Difference between experimental and theoretical (USDA, USDB) binding energies for $N=19$ isotones, $9 \leq Z \leq 17.$  The error bars on the data points represent experimental errors only.  The blue dotted, red dashed, and black dash-dotted bands represent the respective $170$ and $130$ keV RMS deviations of USDA and USDB interactions. Experimental values, save for $Z=9$ which is from the present work, are taken from \cite{Jurado200743} if reported there; otherwise they are from the 2003 Atomic Mass Evaluation \cite{Audi2003337}.}
	\label{fig:n19be}
\end{figure}

Combining the present measurement of the \nuc{28}{F} neutron separation energy with the mass measurements of \cite{Jurado200743}, which place the \nuc{27}{F} atomic mass excess at $246\underline{3}0(1\underline{9}0)$ keV, we determine the \nuc{28}{F} binding energy to be $1860\underline{4}0 (2\underline{0}0)$ keV. As presented in \cite{PhysRevC.74.034315}, it is possible to deduce the presence of ground state intruder components in $N \leq 20$ nuclei by comparing experimental binding energies to $sd$ shell model predictions, such as those of the USDA and USDB effective interactions. For a given nucleus, good agreement between experiment and USDA/USDB theory indicates a ground state configuration that is primarily $sd$ shell.  In contrast, a nucleus with significant ground state intruder components will be poorly described by the USDA/USDB shell model.

\figref{fig:n19be} presents a plot of $\textit{BE}_\textit{exp} - \textit{BE}_\textit{th}$ for $N=19$ isotones, $9 \leq Z \leq 17,$ with the fluorine data point taken from the present work.  The agreement between experiment and USDA/USDB predictions is good for the isotones closer to stability $(Z \geq 13).$  For $Z=10\text{--}12,$ the USDA/USDB calculations predict significantly lower binding than experiment, as expected for these island of inversion nuclei.  For \nuc{28}{F}, the good agreement between experiment and USDA/USDB is recovered, providing evidence that intruder configurations are not significant in the ground state of \nuc{28}{F.}  

In conclusion, we have determined, for the first time, the ground state of \nuc{28}{F} to be a resonance $2\underline{2}0 (\underline{5}0)$ keV above the ground state of \nuc{27}{F} using the technique of invariant mass spectroscopy.  Combined with the mass measurements of \cite{Jurado200743}, this translates to a \nuc{28}{F} binding energy of $1860\underline{4}0 (2\underline{0}0)$ keV.  Investigation of $N=19$ binding energy systematics, including the present measurement, shows good agreement between experiment and USDA/USDB predictions for \nuc{28}{F}, in sharp contrast to the island of inversion nuclei \nuc{29}{Ne}, \nuc{30}{Na}, and \nuc{31}{Mg}.  This indicates that $pf$ shell intruder components play only a small role in the ground state structure of \nuc{28}{F}, establishing a  ``southern shore'' of the island of inversion.

\begin{acknowledgments}

The authors acknowledge the efforts of NSCL operations staff for providing a high quality beam throughout the experiment and the NSCL design staff for their efforts in the construction of a magnetic shield for CAESAR. Additionally, we acknowledge the contribution of the entire MoNA collaboration, past and present, and the NSCL Gamma group for their efforts in preparing and running the experiment.  On the theory side, B. A. Brown and A. Signoracci provided assistance and enlightening discussions regarding shell model calculations.  This work was supported by the National Science Foundation under grants PHY-05-55488, PHY-05-55439, PHY-06-51627, PHY-06-06007, PHY-08-55456, and PHY-09-69173.

\end{acknowledgments}

\end{document}